\documentclass[conference]{IEEEtran}
\IEEEoverridecommandlockouts
\usepackage{cite}
\usepackage{amsmath,amssymb,amsfonts}
\usepackage{algorithmic}
\usepackage{graphicx}
\usepackage{subfig}
\usepackage{textcomp}
\usepackage{xcolor}
\usepackage{hyperref}
\def\BibTeX{{\rm B\kern-.05em{\sc i\kern-.025em b}\kern-.08em
    T\kern-.1667em\lower.7ex\hbox{E}\kern-.125emX}}

\begin{document}

\title{ENCOVIZ: An open-source, secure and multi-role energy consumption
visualisation platform\\
% \thanks{}
}
\author{
    \IEEEauthorblockN{Efstratios Voulgaris\IEEEauthorrefmark{1}, Ilias Dimitriadis\IEEEauthorrefmark{1}, Dimitrios P. Giakatos\IEEEauthorrefmark{1}, \\Athena Vakali\IEEEauthorrefmark{1}, Athanasios Papakonstantinou\IEEEauthorrefmark{2}, Dimitris Chatzigiannis\IEEEauthorrefmark{2}}
    \IEEEauthorblockA{\IEEEauthorrefmark{1}Department of Informatics, Aristotle University of Thessaloniki, Greece
    \\\{voulefst, idimitriad, dgiakatos, avakali\}@csd.auth.gr}
    \IEEEauthorblockA{\IEEEauthorrefmark{2}Energy Management Department, Heron Energy S.A. 
    \\\{apapakonstantinou, dchatzigiannis\}@heron.gr}
}

\maketitle

\begin{abstract}
The need for a more energy efficient future is now more evident than ever and has led to the continuous growth of sectors with greater potential for energy savings, such as smart buildings, energy consumption meters, etc. The large volume of energy related data produced is a huge advantage but, at the same time, it creates a new problem; The need to structure, organize and efficiently present this meaningful information. In this context, we present the ENCOVIZ platform, a multi-role, extensible, secure, energy consumption visualization platform with built-in analytics. ENCOVIZ has been built in accordance with the best visualisation practices, on top of open source technologies and includes (i) multi-role functionalities, (ii) the automated ingestion of energy consumption data and (iii) proper visualisations and information to support effective decision making both for energy providers and consumers.
\end{abstract}


\section{Introduction}\label{intro}
During the past years the number of initiatives tailored to address the constantly aiming higher energy efficiency targets, has increased rapidly\cite{b1}. Especially in the European Union, which has announced a goal of reducing the energy consumption by, at least, 32.5\%  by 2030, there is an ongoing effort to engage citizens in adopting a more energy efficient attitude \cite{b2}. Many of these efforts, leverage the use of multiple types of energy monitoring devices and sensors, which in turn produce a really large volume of exploitable energy related data. Such approaches can be combined with dashboards, to provide near real-time energy usage feedback and encourage individuals to comprehend and effectively regulate their energy usage. 

A dashboard can be considered as a decision support system driven by data that delivers information to the decision maker in a specific format\cite{b3}. The number of commercial energy monitoring platforms, including energy dashboards, is steadily growing, a clear indication of the popularity of such solutions. At the same time, similar platforms have also attracted the interest of researchers for more than a decade \cite{b4,b5,b6}, who pointed out that such solutions can play an important role in the effective reduction of energy consumption. In particular, a recent research \cite{b7} highlighted the fact that a group of people who were provided with access to an intelligent energy dashboard along with specific interventions, achieved much higher energy savings in their workplace in comparison to others who didn't have access to any of these resources. 

On the other hand, according to some researchers, the use of dashboards for energy management and monitoring energy consumption is not necessarily linked to a corresponding reduction in consumption \cite{b8,b9,b10}. More specifically, they have shown that the use of such tools should be accompanied by other interventions at both a societal and personal level in order to be effective. This finding is in accordance with the experiment mentioned above\cite{b7}.

Of course, the development of an effective energy management platform involves many challenges, which, as seen from the literature, are a common finding among most researchers. Although energy management platforms may appear relatively uncomplicated, they do pose extensive technical challenges both at architectural and visualisation design levels\cite{b11}. Taking into consideration both of these aspects, the first challenge is identifying the most effective visualization method, while the second challenge is integrating the larger information system in a way that empowers residents to make informed decisions. The key to enabling consumers, who may not be technology experts, to understand their energy consumption is developing these information systems with simplicity in mind\cite{b5}. 

\begin{figure*}[ht]
\centering
\includegraphics[width=1\textwidth]{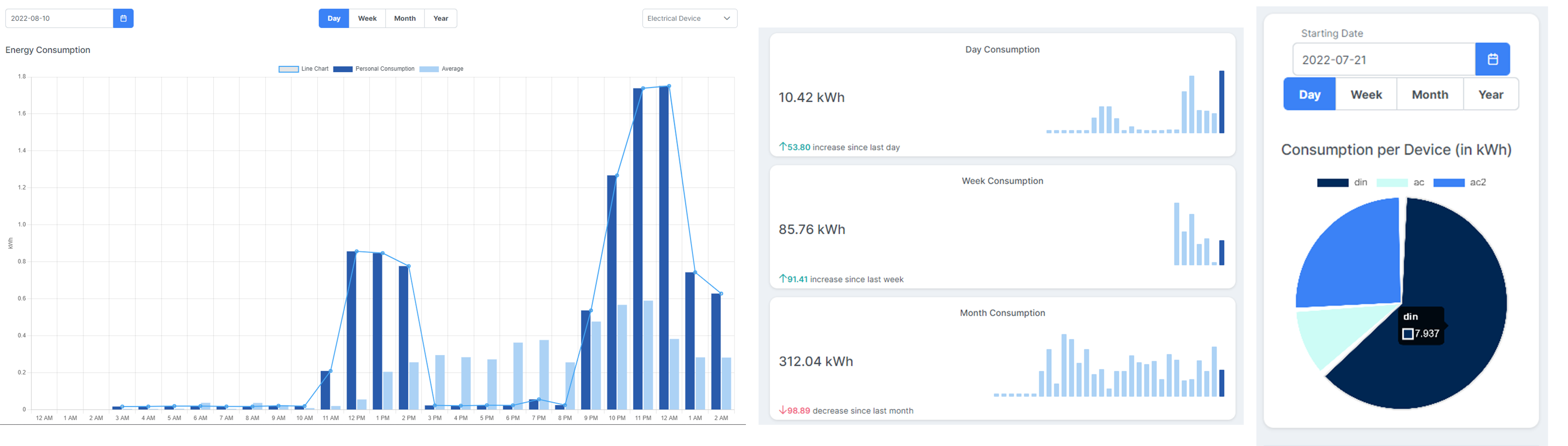}
\caption{ENCOVIZ consumers' energy visualisations}
\label{fig1}
\end{figure*}

Moreover, the development of such platforms includes a wide variety of requirements to be met, such as: \textbf{[R1]}: Multi-role access and role-based content delivery (energy provider, consumer, admin), \textbf{[R2]}: Data Security and encryption, \textbf{[R3]}: Automated ingestion, processing and analysis of data and \textbf{[R4]}: Responsive design, clear, precise and non-redundant visualisations, which conform with state-of-the-art dashboard visualisation principles \cite{b12,b13,b14}. On top of that, taking into consideration the lack of publicly available solutions, the use of open source, long-term supported and well-tested technologies is definitely beneficiary \textbf{[R5]}.

In this sense, this paper presents ENCOVIZ a web platform for the visualisation and analysis of residential energy consumption measurements, see Figure~\ref{fig1}, which addresses all the aforementioned challenges and requirements. It has been developed and used for a short trial period by a group of Greek home residents in the framework of the Heart 
 project\footnote{\url{http://heartproject.gr/}}. ENCOVIZ allows energy providers to easily set up an energy management platform, which monitors the energy consumption of end-users and provides visual aids towards effective energy reduction decisions.

 The rest of this paper is structured as follows: Section \ref{data_sources} presents the data processing methodology followed in the proposed solution. Section \ref{architecture} describes the overall architecture of the platform, Section \ref{features} introduces the main features of ENCOVIZ and the results of the platform's evaluation by real users. Finally, Section \ref{conclusion} concludes this paper and indicates future research directions.

\section{Data Processing}\label{data_sources}
\begin{figure}[h]
\centerline{\includegraphics[width=0.35\textwidth]{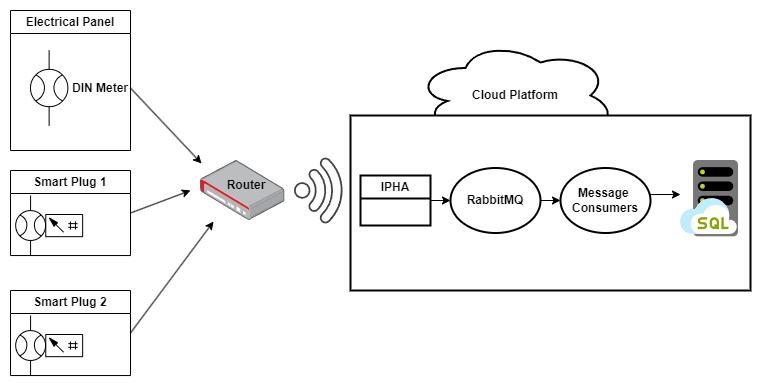}}
\caption{Data Ingestion Pipeline. Smart plugs and the DIN meter collect the energy consumption of two devices and the total energy consumption of each home accordingly, which are then forwarded to the cloud}
\label{fig2}
\end{figure}

 The present work has been based on data that has been collected within the Heart project, which employs IoT and smart meters to collect energy consumption data of home residencies and devices for a period of three months. More specifically, two smart plugs and one DIN meter, which calculates the total energy consumption of a home, have been installed in thirty homes. This data are then forwarded to a data collection platform hosted in the cloud. Both the data ingestion pipeline and the cloud database have been developed by Net2Grid\footnote{\url{https://www.net2grid.com/}} (Heart Project partner) for the purposes of the project, see Figure~\ref{fig2}, but are beyond the scope of this paper. The data stored in the cloud platform include the following information: 
 \begin{itemize}
     \item Timestamp: Time of measurement in UNIX ms timestamp format (measurement frequency 1Hz)
     \item Value: The corresponding value for either the total consumption or the device consumption in Watts
     \item deviceID: Unique identifier for each of the metric devices (DIN, Smart plugs)
 \end{itemize}
 
As mentioned, each house has two smart plugs installed on specific devices (Washing machine, dryer, dishwasher, A/C units, TV, etc.) and one DIN meter. These three meters export data with the energy consumption in Watts and the timestamp of the measurement in UNIX milliseconds format. This data are exported into CSV files, one for each device, where the file name corresponds to the device's unique ID. The CSVs are exported by the cloud component and are then uploaded into the ENCOVIZ system in order to be processed. Although this manual process of getting the data and uploading them to the system might not be optimal, ENCOVIZ also supports an API in which data can be imported directly, without generating any CSV files. Still, we leave this change as a future work. After the data has been uploaded to ENCOVIZ, the system administrator should use the dashboard to start the data syncing. During the syncing, the system runs two processes; one for updating the users' data and one for updating the provider's data.

\textbf{Updating users' data.} This process employs a user-device mapping, available in the system's database, in which someone can find the devices' unique ID for each user. This mapping can be considered as an initial configuration file, which the provider has to create for the whole process to be completed. The data syncing process includes queries of specific device IDs to the mapping file to retrieve the exported files accordingly. The data are then sorted according to the timestamp for each day, and aggregated to the energy consumption in hourly format. Finally, the data are imported into the ENCOVIZ database.
% When the process starts the data syncing from the CSV files, it searches in the hardware dictionary for each file name the user that has in his house this device.

\textbf{Updating providers’ data.} After the completion of the first process, the updated data are used as input to the second process, which syncs the provider's data. This process sums the hourly energy consumption per day from all users and stores them to the ENCOVIZ database along with the total number of the current users.
% After the import of all the new data, the system  this data into the second process in order to sync the provider’s data. This algorithm sums the hourly energy consumption per day from all users and imports them into ENCOVIZ database along with the total number of the current users.
The whole syncing operation is created using Redis for controlling multiple requests from providers. Although in this use case, we have only one provider, ENCOVIZ can already support many. In this way, when multiple providers are going to sync their customers' data the system will not fail due to low hardware capabilities from the hosting machine. As a database, we use MongoDB due to the NoSQL structure of the data that we are processing. Redis and MongoDB transactions are protected using Docker, as they are installed in separate virtual machines inside the server without exposing their IPs and ports to the Internet. Furthermore, ENCOVIZ API is protected using Keycloak, but it is not limited to. It can support many other authentication technologies.

\textbf{Analytics.} When users or providers make a request to view their data, the system makes aggregations according to the period and the time unit that they have selected. The system can display the energy consumption per hour, week, month, and year and returns the consumption in a specific date range. It also calculates the average consumption of all users and devices for a specific selected period and the differences between previous time periods, as presented in Figure~\ref{fig1}.
\section{Encoviz design and architecture}\label{architecture}
\begin{figure}[h]
\centerline{\includegraphics[width=0.45\textwidth]{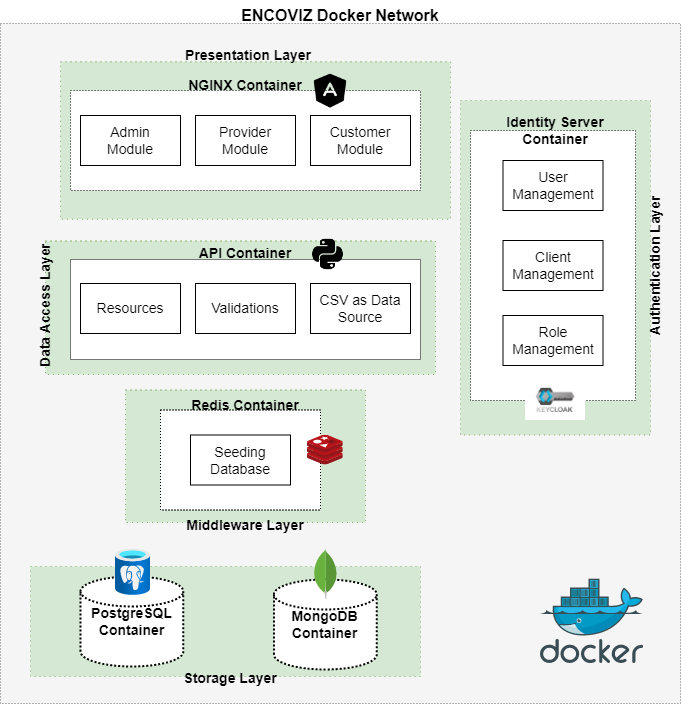}}
\caption{Architecture Layers of the ENCOVIZ platform}
\label{fig3}
\end{figure}

Developing an effective energy management platform requires careful consideration of both the requirements set by experts in the field and the requirements related to software development. ENCOVIZ has been designed to provide comprehensive monitoring capabilities, allowing users to analyze energy usage across a range of applications. Functional requirements such as data ingestion and analysis must be carefully integrated with non-functional requirements such as system performance, scalability, and security. With that in mind and based on the requirements described at Section\ref{intro} ENCOVIZ has been organized in five different layers, each of which consists of individual components, see Figure~\ref{fig3}. 

\begin{figure}[h]
\centerline{\includegraphics[width=0.5\textwidth]{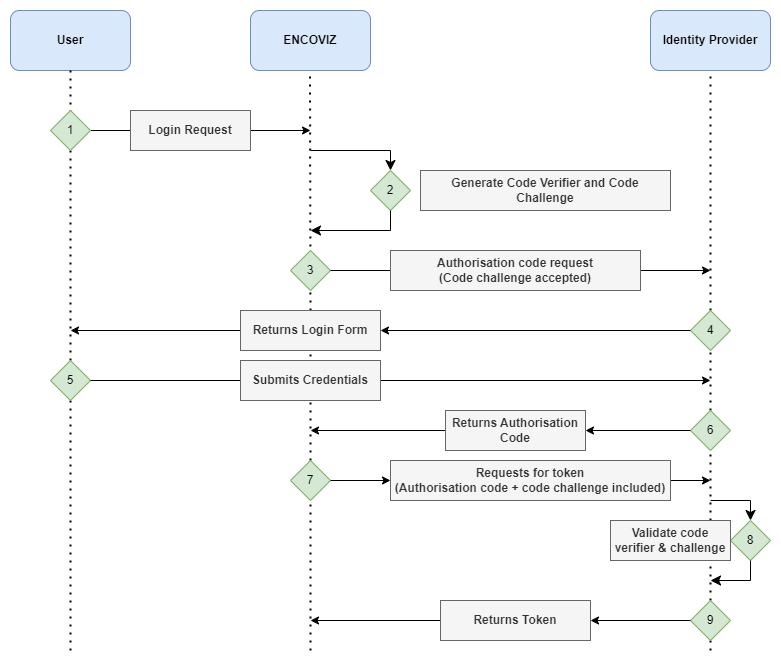}}
\caption{Authentication process of the ENCOVIZ platform}
\label{fig4}
\end{figure}

\textbf{Authentication Layer [R1,R2]:} ENCOVIZ serves both the needs of consumers and  energy providers. For this reason and since we handle sensitive data, we follow a role-based access control (RBAC) approach, where we establish roles for defining and restricting the level of access of users to the various resources of our system. At a very high level, we need a service/provider that can verify user’s credentials and allow or deny access on the system. We use the OpenID Connect, which is an open authentication protocol that works on top of OAuth 2.0 framework, and Keycloak\footnote{\url{https://www.keycloak.org/}} as an Identity provider and Authentication Server. The role of the provider is to give, via the selected authentication flow, the ability to the client or service (called Relying Party) requesting a user’s identity to log in. There are different flows that can be defined in the OpenID protocol depending on the type of application and its security requirements. The differences of these flows have mainly to do with what response types an authorization request can request and how tokens are returned to the client application. We will use Authorization Code Flow with Proof Key for Code Exchange (PKCE). This flow requires the least amount of effort for the client application to implement while giving the best level of security but most importantly is recommended as best practice from OAuth2\footnote{https://oauth.net/2/} foundation especially for single page applications, as ENCOVIZ. The authentication process is further described in Figure~\ref{fig4}.
Both the ID and access token are encoded as a JSON Web Token (JWT), a standard format that enables the client application to inspect its content and verify that it originates from the expected issuer and hasn’t been altered by anybody else. The ID token might include information regarding the identity of the logged in user (name, email, username, profile picture, etc.) but is mainly used to indicate that the user has been authenticated by the OpenID provider. The access token defines which resources can be accessed by the user that has successfully been authenticated. The communication within the Identity server and the application is achieved in a REST-like manner.
% Keycloak includes a REST API that allows us to perform many actions programmatically, provides a Graphical UIO for every needed configuration and it is sponsored by Red Hat, ensuring that support and maintenance are not an issue. 

\textbf{Presentation Layer [R4]:} The UI has been developed using Angular, which supports routing, client-server communication and a wide range of 3rd party libraries. We use the PrimeNg\footnote{\url{https://primeng.org/}} library which provides out-of-the-shelf fully responsive components. After a user successfully signs in, the authorization server issues an access token indicating which resources are permitted to be accessed from the client application. The views are divided into three separate modules, one for each user role. For this reason we developed a mechanism that determines which module should be loaded based on the received access token when the application is initially instantiated. Since only the required content is loaded, the application becomes more lightweight in terms of bundles’ size. The type of visualisations included are further described in section~\ref{features}\\*
\textbf{Storage Layer[R2]:} This layer consists of two database components. The first, PostgreSQL\footnote{\url{https://www.postgresql.org/}}, is totally independent from the application as it contains data that are related to the users of the system (role, username, password, etc.), authentication flows, security settings etc. The second one, MongoDB\footnote{\url{https://www.mongodb.com/}}, contains all the data related to the dashboard app and especially the information which will be displayed, like energy consumption (daily, weekly, monthly), user’s devices, energy consumption per device, etc. At this point we should also highlight that the choice of using two different databases was defined by the project's specifications. Instead we could either use one SQL database for both the users and the energy data, or a noSQL database encapsulated via a CRUD service to improve exchangeability.\\* 
\textbf{Data Access \& Messaging Layer [R3]:} The Data Access Layer consists of an API that is responsible for the delivery, insertion and processing of raw data, as presented in Section~\ref{data_sources}. The data in our case are in a CSV format and each CSV file initially corresponds to energy consumption measurements. The API relies on three different components: FastAPI\footnote{\url{https://fastapi.tiangolo.com/}}, MongoDB, and Redis\footnote{\url{https://redis.io/}} which acts as a middleware (message broker). The API endpoints allow the user to perform queries to the database and get information about the energy consumption per certain period and per device, as well as other statistics (average consumption, min, max, etc.). For the insertion process we use Redis, where we store the paths of the files that contain the users’ electric consumption. The data that will then be visualised, e.g. average consumption of a device, is available by making GET requests to certain endpoints. \\*
As we can see the whole framework makes use of mature, open-source technologies, thus successfully complies with \textbf{R5}. Our framework has been containerized using Docker\footnote{\url{https://www.docker.com/}}. We open source our code and Docker image at \url{https://github.com/Datalab-AUTH/encoviz}, where we also provide detailed installation steps. 

\section{Encoviz features and evaluation}\label{features}
In this section we are going to present the main features and visualisations of our platform. The platform can be accessed by three different types (roles) of users: (a) Consumers, (b) Energy providers, (c) Administrators. Although there are some features that are applicable to everyone, each user has different expectations from the platform depending on their role. These expectations define the features and visualisations. The selected visualisations, defined based on state-of-the-art guidelines\cite{b3,b12} and the experts opinion, were bar charts, line charts and combinations between them for visualising the distribution, comparison and trend of energy consumption, while pie charts were selected for the composition of energy consumption per device. The filters used to receive this information include date filtering, time unit filtering(Day, Week, Month, Year), electrical device and consumers' IDs, depending on each use case. All visualisation conform with the rule of perceiving the maximum amount of data in a minimum amount of time, while remaining easy to use and comprehend, see Figure~\ref{fig1}. The main features of our platform can be categorized as follows:

\textbf{Global features:} Through Keycloak's Account Management Console, each user has the ability to (i) edit their personal info, (ii) track their activity, (iii) change their password. \\*
\indent\textbf{Administrative features: }The platform provides access to admins who can see a list containing all users along with some of their corresponding data. This data can prompt administrators to take certain actions, such as reminding users to verify their emails.\\*
\indent\textbf{Consumer features:} The system provides: (i) a brief overview of their current daily, weekly, monthly total and per device consumption, (ii) comparison with the average consumption of other users, (iii) facilitates users to view their consumption in accordance with various criteria that themselves can specify and (iv) provides a view of their electrical devices which includes a list of their devices along with their individual consumption.  \\*  
\indent\textbf{Provider's features:} Providers are able to see the energy consumption making every possible combination throughout the available filters. They can also see every device category that is inserted in the system and the total consumption for each one of the devices and users. These users access the full range of system capabilities since they often combine data from several sources in their daily operations.

\textbf{ENCOVIZ evaluation}
The proposed platform has been successfully tested on the Heart research project by more than 30 individuals consumers and an energy provider company. Following the conclusion of the testing period, interviews were conducted with 10 users. The interview focused on the installation process and general equipment issues such as connectivity, and assessed the user interface and its relevance as a tool that would ultimately assist in reducing consumption. 100$\%$ of the users were positive (Very Satisfied and Satisfied) with the user registration following the installation from an electrician. 20$\%$ were positive, 70$\%$ indifferent and 10$\%$ negative on the connectivity issues and the effort it took on their behalf to re-connect to their devices when experiencing connectivity issues. Out of the 70$\%$ which was indifferent, all of them said they did not experience any issues, hence their attitude. Regarding the user interface, 80$\%$ were positive in terms of the presentation, however when this group was asked whether they intended to reduce their consumption, only half said they would, with the other half being indifferent. Those who were inclined to modify their behaviour, said they were motivated by the collective behaviour of the community (comparison with average). This conclusion is in accordance with other studies, that promote the term of "energy-voyeurism"\cite{b10} and is an interesting finding. Now, the rest 20$\%$ on the user interface, was indifferent, citing complexity and overall lack of interest. When asked whether they have noticed the comparison features, both replied negatively, mentioning that a more positive attitude could have been triggered, had they realised the social aspects.
% In this section we summarize the overall effort and highlight future research directions based on the initial feedback received by users of the platform. 
\section{Conclusions and Future Work}\label{conclusion}
In this paper we present ENCOVIZ a secure open source web visualisation platform for energy consumption data visualisation, which attempts to address open challenges with respect to other similar solutions. It supports multi-role functionalities, state-of-the-art visualisation practices and offers data handling by default. Although it has been presented as an energy management tool, ENCOVIZ can be used in other similar areas, such as water management or other resource management tasks. 

The short feedback that has been received during the pilot phase, where an initial set of users experimented with the proposed platform indicate points that need further improvement, but at the same time showcase the potential of ENCOVIZ. Future work includes further optimisation of the user interface, instant connection to APIs for data retrieval, direct integration of NILM models and the addition of more features that would highlight the social aspect of the proposed solution.

\section*{Acknowledgment}
This research is co-financed by Greece and European Union through the Operational Program Competitiveness, Entrepreneurship and Innovation under the call RESEARCH-CREATE-INNOVATE (project T2EDK-03898)
% \section*{References}

% Please number citations consecutively within brackets \cite{b1}. The 
% sentence punctuation follows the bracket \cite{b2}. Refer simply to the reference 
% number, as in \cite{b3}---do not use ``Ref. \cite{b3}'' or ``reference \cite{b3}'' except at 
% the beginning of a sentence: ``Reference \cite{b3} was the first $\ldots$''

% Number footnotes separately in superscripts. Place the actual footnote at 
% the bottom of the column in which it was cited. Do not put footnotes in the 
% abstract or reference list. Use letters for table footnotes.

% Unless there are six authors or more give all authors' names; do not use 
% ``et al.''. Papers that have not been published, even if they have been 
% submitted for publication, should be cited as ``unpublished'' \cite{b4}. Papers 
% that have been accepted for publication should be cited as ``in press'' \cite{b5}. 
% Capitalize only the first word in a paper title, except for proper nouns and 
% element symbols.

% For papers published in translation journals, please give the English 
% citation first, followed by the original foreign-language citation \cite{b6}.

\vspace{12pt}
% \color{red}
% IEEE conference templates contain guidance text for composing and formatting conference papers. Please ensure that all template text is removed from your conference paper prior to submission to the conference. Failure to remove the template text from your paper may result in your paper not being published.
\end{document}